\documentclass[10pt,conference]{IEEEtran}
\usepackage[utf8]{inputenc}
\usepackage{cite}
\usepackage{algorithm}
\usepackage{amsmath, amssymb}
\usepackage{algpseudocode,algorithm,algorithmicx}
\usepackage{subfiles}
\usepackage{url}
\usepackage{graphicx}
\usepackage{caption}
\usepackage{subcaption}
\usepackage{textcomp}
\usepackage{bm}
\usepackage{xcolor}
\usepackage{amsthm}

\usepackage{changes}
\definechangesauthor[name="Ilya Safro", color=red]{is}
\definechangesauthor[name="Ankit", color=purple]{ak}

\title{BEINIT: Avoiding Barren Plateaus in Variational Quantum Algorithms}
\author{\IEEEauthorblockN{Ankit Kulshrestha}
\IEEEauthorblockA{\textit{Computer and Information Sciences} \\
\textit{University of Delaware}\\
Newark DE, USA \\
akulshr@udel.edu}
\and
\IEEEauthorblockN{Ilya Safro}
\IEEEauthorblockA{\textit{Computer and Information Sciences} \\
\textit{University of Delaware}\\
Newark DE, USA\\
isafro@udel.edu}}

\DeclareMathOperator*{\argmin}{arg\,min}

\newcommand{\beinit}{BEINIT }
\newcommand{\ket}[1]{\ensuremath{|#1 \rangle}}
\newcommand{\bra}[1]{\ensuremath{\langle #1 |}}

\newcommand{\U}{U(\vec{\theta})}
\newcommand{\Uad}{U^{\dagger}(\vec{\theta})}

\newtheorem{conj}{Conjecture}

\date{March 2022}

\begin{document}

\maketitle
\begin{abstract}
Barren plateaus are a notorious problem in the optimization of variational quantum algorithms and pose a critical obstacle in the quest for more efficient quantum machine learning algorithms. Many potential reasons for barren plateaus have been identified  but few solutions have been proposed to avoid them in practice. Existing solutions are mainly focused on the initialization of unitary gate parameters without taking into account the changes induced by input data. In this paper, we propose an alternative strategy which initializes the parameters of a unitary gate by drawing from a beta distribution. The hyperparameters of the beta distribution are estimated from the data. To further prevent barren plateau during training we add a novel perturbation at every gradient descent step. Taking these ideas together, we empirically show that our proposed framework significantly reduces the possibility of a complex quantum neural network getting stuck in a barren plateau.\\
\noindent Reproducibility: The source code and data are available at \url{https://github.com/aicaffeinelife/BEINIT} 

    
    \end{abstract}

\section{Introduction}
The world of machine learning has seen a tectonic paradigm shift in the last decade. From handcrafted features that were manually fed to classical optimization algorithms like support vector machines we have moved on to the deep neural networks (DNN) that can simultaneously perform the task of feature engineering and recognition without any explicit intervention from the developer. The ever increasing performance gain in terms of accuracy on really large datasets has led to the quest for more faster and accurate algorithms that can scale to the amount of data being produced today.
 
 The ongoing development of Noisy Intermediate-Scale Quantum (NISQ) computers has led to considerable excitement about the potential quantum advantage that can be obtained in several optimization problems that are used in almost all fields of science today. Various hybrid quantum-classical algorithms \cite{qaoa20142,cerezo2021variational,ushijima2021multilevel,shaydulin2019hybrid} offer possible ways to utilize  NISQ computers. Variational Quantum Algorithms (VQAs) offer a potential way of leveraging quantum computers alongside classical computers in a hybrid fashion. When it comes to deployment of VQAs on NISQ devices, the limitations of existing quantum computers (i.e., noisy gates and limited circuit depth) \cite{cerezo2021variational,liu2022layer} restrict the overall potential of these algorithms. Moreover, VQAs themselves are not without issues. In this paper, we make a step towards addressing a well known issue that frequently arises in optimization approaches involving VQAs.
 
 The main component of a VQA is a variational quantum circuit (VQC) that consists of a finite sequence of paramaterized unitary gates $\U$ and a measurement produced by a quantum observable $\hat{O}$. The parameters of the unitary gates are real-valued. More formally,  we can express $\U$ as:
 
 \begin{equation}
     \U = \prod_{i=1}^{L} U(\theta_{1})U(\theta_{2})\dots U(\theta_{L})
     \label{eq:vqc}
 \end{equation}
 
 An ansatz formed of a sequence of unitary gates as in Equation~\ref{eq:vqc} is trained to solve an optimization problem of the form:
 \begin{equation}
     \theta^{*} = \argmin_{\theta} C(\theta)
     \label{eq:opt-prob}
 \end{equation}
 
 Where, $C(\theta)$ is a \emph{cost function} that quantifies the deviation of a VQC at a given timestep from the ``true" trajectory of the problem. Given $\hat{O}$ and $\theta$ a general cost function for VQAs can be defined as:
 
 \begin{equation}
     C(\theta) = f(\bra{0}\Uad \hat{O}\U \ket{0})
     \label{eq:cost-fn}
 \end{equation}
 
 Where the choice of $f$ depends on the specific application of the VQA~\cite{cerezo2021variational}. The parameters of a VQC are initialized from a random distribution. Over the course of training, the output of the VQC is measured with respect to the observable $\hat{O}$ and the gradients of every unitary gate are estimated using the parameter shift rule~\cite{schuld2019evaluating, mitarai2018quantum}. The set of parameters at a given time step and the gradients of the unitary gates are then passed to a classical computer which updates these parameters according to a gradient descent rule. This update is usually performed by an \emph{optimizer} and it's choice is usually considered to be a hyperparameter(i.e. parameters which are manually selected by the developer). Some popular optimizers include Adam~\cite{kingma2014adam}, AdaGrad~\cite{duchi2011adaptive} etc.



Ideally, for an optimization problem, the set $\theta^{*}$ should correspond to the best set of parameters that minimizes the cost function. However, it has been observed that when the VQC is complex either in number of qubits used to represent the input or the number of layers of unitaries, the optimization halts at a suboptimal set of parameters that do not correspond to a minima in the optimization surface. This situation occurs when the circuit gets stuck on a plateau from  where there are no good descent directions and is commonly referred to as being stuck in a ``barren-plateau". The unitary designs induced due to random parametrization of ansatz~\cite{mcclean2018barren}, optimization w.r.t a global cost function~\cite{cerezo2020cost2}, entanglement between visible and hidden layers of a Quantum Neural Network (QNN)~\cite{marrero2021entanglement} and random entanglement~\cite{patti2021entanglement} have been studied as potential causes for the barren plateau problem. Attempts to mitigate the barren plateau problem comprise of advanced initialization schemes~\cite{grant2019initialization, verdon2019learning2}, local objective functions~\cite{cerezo2020cost2} and Bayesian initialization schemes~\cite{rad22bayesian}.\\

\emph{Our contribution} The current body of work leaves, among the others, two questions unanswered. First, does the choice of the initializing distribution have an effect on the appearance of barren plateaus in a random ansatz (barring the case where the parameters are initialized from a classical neural network)? Second, once a good initialization has been found, is there a way to influence the gradient descent such that the likelihood of barren plateaus is reduced? In this paper, we study these questions for the specific case of QNNs and empirically show that the choice of initializing distribution is an important and often neglected hyperparameter in VQAs. We further show that adding a derived perturbation in the parameter space during gradient descent can help a QNN from getting stuck in barren plateaus. We combine these insights into an algorithm which we call \beinit. Although our results are derived for the specific case of QNN, we believe that these results are general enough for any VQA application.


\begin{figure*}[t!]
\centering
\begin{subfigure}[t]{.55\columnwidth}
    \centering
    \includegraphics[height=1.5in, width=1.5\textwidth]{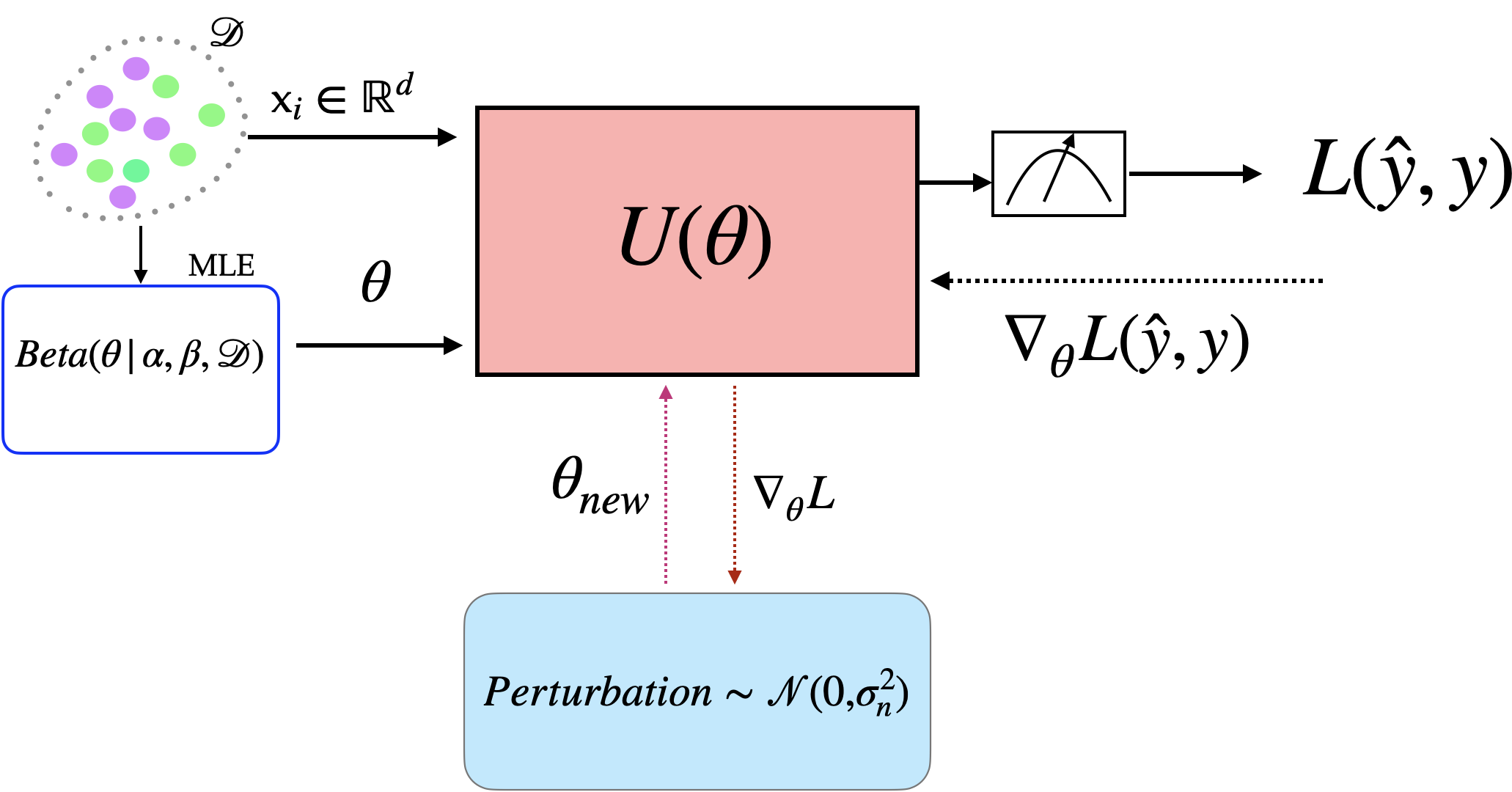}
    \caption{Overview of the \beinit algorithm}
    \label{sfig:beinit}
\end{subfigure}
\hspace{1.5in}
\begin{subfigure}[t]{.55\columnwidth}
    \centering
    \includegraphics[height=1.75in, width=1.45\textwidth]{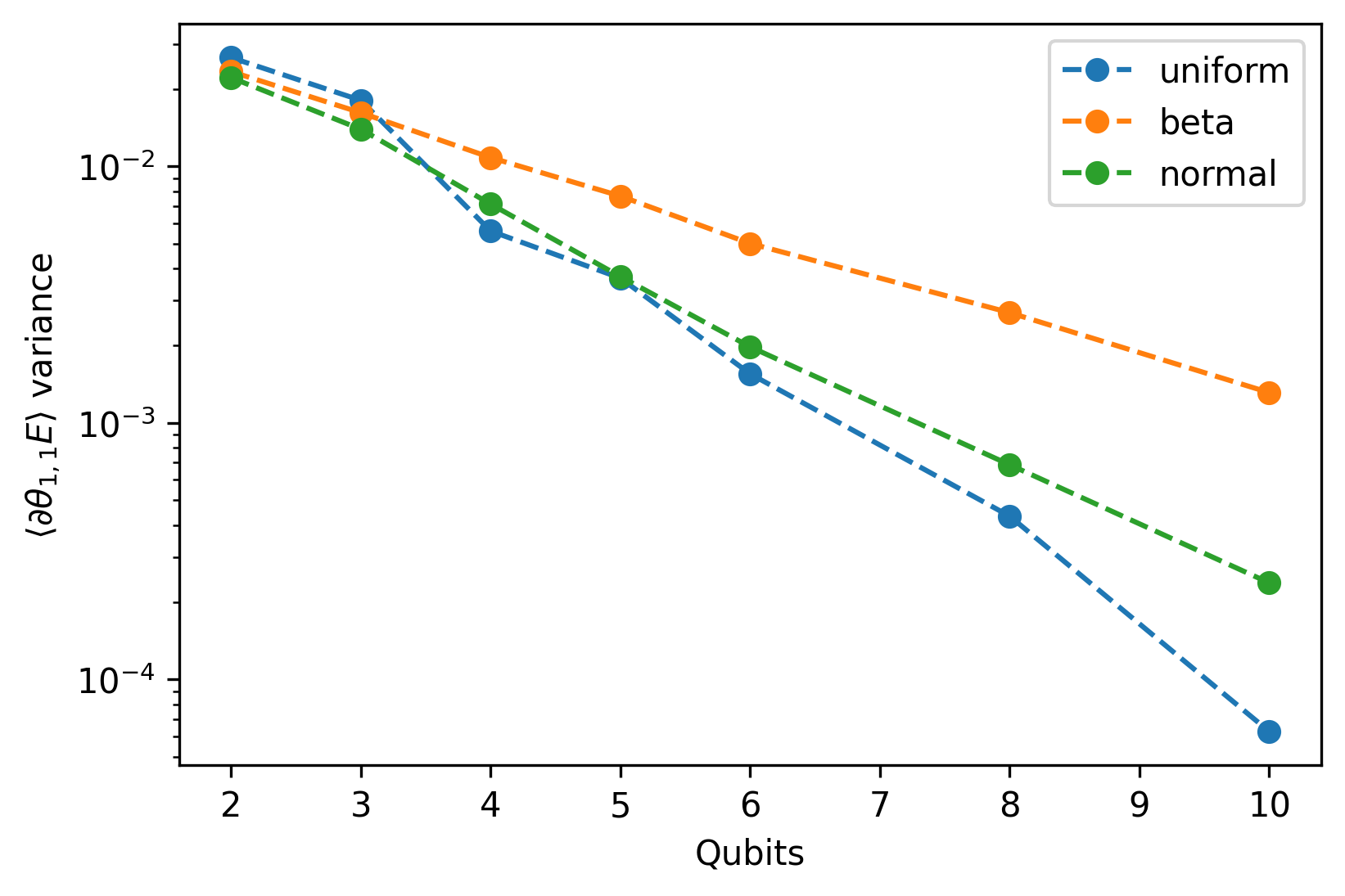}
    \caption{Variance of Gradient with VQC initialized by different distributions.}
    \label{sfig:beta-expt}
\end{subfigure}

\caption{The \beinit algorithm accepts data and parameters generated from a beta distribution whose hyperparameters are generated from an MLE estimation of the data. The gradients are then utilized to generate a perturbation in the parameter space. The empirical basis for the algorithm is shown in the right figure.}
\label{fig:beinit_alg}
\end{figure*}



\section{Barren Plateau Problem in QNNs}
We earlier mentioned that barren plateaus can have several causes. In this work, we choose to focus on a generic case where we only vary the parameters of the unitary gates (as opposed to additionally randomizing the specific gate choice as in~\cite{mcclean2018barren}) and optimize in presence of a global cost function i.e. for a given QNN we have one global observable. The global cost function assumption is not unreasonable since we wish to study the effect of initialization in cases where no additional steps to mitigate barren plateaus have already been taken (e.g. optimizing with local observables~\cite{cerezo2020cost2}). 


Since we deal with QNNs, we briefly specify the general VQA framework discussed earlier. For any labeled dataset consisting of $m$ samples; $D = \{(x_{i}, y_{i})\}_{i=1}^{m}$, a QNN requires access to a state encoding function $K(\vec{x})$ which accepts $\vec{x}^{(i)} \in \mathbb{R}^{d}$ and produces a quantum state. This encoding can be seen as an instance of kernel function which maps a real valued vector into the complex Hilbert state. The output state before measurement of a QNN can be described as:

\begin{equation}
    \ket{\psi} = \U K(\vec{x}) \ket{0}.
    \label{eq:qnn-out}
\end{equation}

\noindent
The output of a classifier $f(\vec{x}, \vec{\theta}, \hat{O})$ is an expectation over $m$ samples in the given dataset:

\begin{equation}
    f(\vec{x}, \vec{\theta}, \hat{O}) = \sum_{i=1}^{m}||y_{i} - \bra{0}\Uad K^{\dagger}(\vec{x}^{(i)}) \hat{O} \U K(\vec{x}^{(i)})\ket{0}||_{2}.
    \label{eq:vqc-output}
\end{equation}

This output is a scalar value which is compared using a distance measure with the true label and the error signal is propagated to each unitary circuit. An advantage of using such differentiation rule as in \cite{schuld2019evaluating} is that the same unitary circuit can evaluate the output and its respective gradient.  A  circuit with unitary gates of dimension $d$ corresponds to a unitary group $U(d)$. For any measure $dU$ on this group, the $k^{th}$ moments can be expressed as in  \cite{verdon2019}:
\begin{equation}
    M_{k}(dU) = \int_{U(d)} dU U_{i1, j1}\dots U_{ik,jk}U^{\dagger}_{i1', j1'}\dots U^{\dagger}_{ik',jk'}.
    \label{eq:general-kth-moments}
\end{equation}

Where $ik, jk$ are the row and column indices of the $k^{th}$ unitary matrix. If the measure is a left invariant measure like Haar-measure, i.e. $dU = dU_{H}$ then, the first and second order moments are given as 
\[
M_{1}(dU_{H}) = \frac{1}{d} \delta_{i1i1'}\delta_{i2i2'}
\]
and 
\begin{align*}
M_{2}(dU_{H}) = \frac{ (\delta_{i1i1'}\delta_{i2i2'}\delta_{j1j1'}\delta_{j2j2'} + \delta_{i1i2'}\delta_{i2i1'}\delta_{j1j2'}\delta_{j2j1'} )}{d^{2} - 1} -\\
\frac{ (\delta_{i1i1'}\delta_{i2i2'}\delta_{j1j2'}\delta_{j2j1'} + \delta_{i1i2'}\delta_{i2i1'}\delta_{j1j1'}\delta_{j2j2'})}{d(d^{2}-1)},
\end{align*}
respectively. Any measure $dU$ exhibits a unitary 2-design if and only if $M_{1}(dU) = M_{1}(dU_{H})$ and $M_{2}(dU) = M_{2}(dU_{H})$. For a  sequence of parameterized unitary gates expressed as 
\[
\U = \prod_{i=1}^{L} U(\theta_{1})\dots U(\theta_{L}),
\]
we can denote 
\[
U_{-} = \prod_{i=1}^{k} U(\theta_{1})\dots U(\theta_{k-1})
\]
and
\[
U_{+} = \prod_{i=1}^{k} U(\theta_{k})\dots U(\theta_{L})
\]
for the $k^{th}$ unitary $U_{k}$

In~\cite{mcclean2018barren} the authors show that if any $U_{-}$ or $U_{+}$ exhibits a unitary 2-design then the variance of the classifier with respect to $\theta_{k}$, $Var[\partial_{k} f]$ can be given by: 

\begin{equation}
    Var[\partial_{k} f] \propto \left( \frac{Tr(U^{2})}{2^{3n}} - \frac{Tr(U)^{2}}{2^{4n}} \right )
    \label{eq:variance-bp}
\end{equation}

In Equation~\eqref{eq:variance-bp}, $U \in U(d)$ and $n$ is the number of qubits. The relation assumes that both $U_{-}$ and $U_{+}$ exhibit a 2-design and shows that with increasing number of qubits, the variance of the gradient will approach zero and consequently the QNN will be stuck in barren plateau with all conditions kept the same (other cases also exhbit a similar kind of decay).

\section{An Experiment With Initialization}
We noted that related works are relying on the common assumption that the parameters for $l^{th}$ layer unitary $U_{l}(\theta_{l})$ are drawn from a uniform distribution (e.g. from $[-\pi, \pi]$). In order to answer our first question i.e. the effect of initialization on variance of gradient with different probability distributions, we reproduced the experimental setup of~\cite{mcclean2018barren} and ran it with parameters derived from the Uniform ($\theta_{l} \sim Unif(0, 2\pi)$), Beta ($\theta_{l} \sim Beta(1, 2\pi)$) and a normal distributions ($\theta_{l} \sim \mathcal{N}(0, 2\pi)$).The results of our experiment are shown in Figure~\ref{sfig:beta-expt}.

Note that the blue line corresponding to the uniform distribution closely matches the results obtained in~\cite{mcclean2018barren}. Interestingly, when the parameters of the quantum circuit are initialized using the beta distribution (orange line), the reduction in variance of the gradient is significantly less than the other two cases. This result may seem counter-intuitive at first, but if we examine the Beta distribution more closely then we find that its domain is $[0, 1]$ which implies that it can be used to model a \emph{distribution over probabilities}. Moreover, its shape parameters $\alpha, \beta$ (which are inputs to the gamma function $\Gamma(z) = \int_{0}^{\infty} x^{z-1} e^{-x}dx$, see Eq. \ref{eq:alpha-mle} and \ref{eq:beta-mle}) can be tuned and used to guide the relative weight over the input probabilities. This behavior is in contrast with the uniform distribution which places equal weight over the input interval.

Based on our experimental observations above we advocate for the following conjecture.

\begin{conj}
The choice of the sampling distribution over a unitary group $U(d)$ has a direct impact on the likelihood of the resulting VQC exhibiting a unitary-2 design. The more non-mean collapsing behavior a distribution exhibits, the lesser the likelihood of the occurence of a barren plateau.
\end{conj}

By a ``mean-collapsing" distribution we refer to a distribution for which its first and second order moments are unbiased MLE estimators (e.g., normal distribution). Conversely, a non-mean collapsing distribution has biased estimators in either first or second order moments.

\begin{figure*}[t!]
\centering
\begin{subfigure}[b]{.3\textwidth}
\centering
\includegraphics[width=\textwidth]{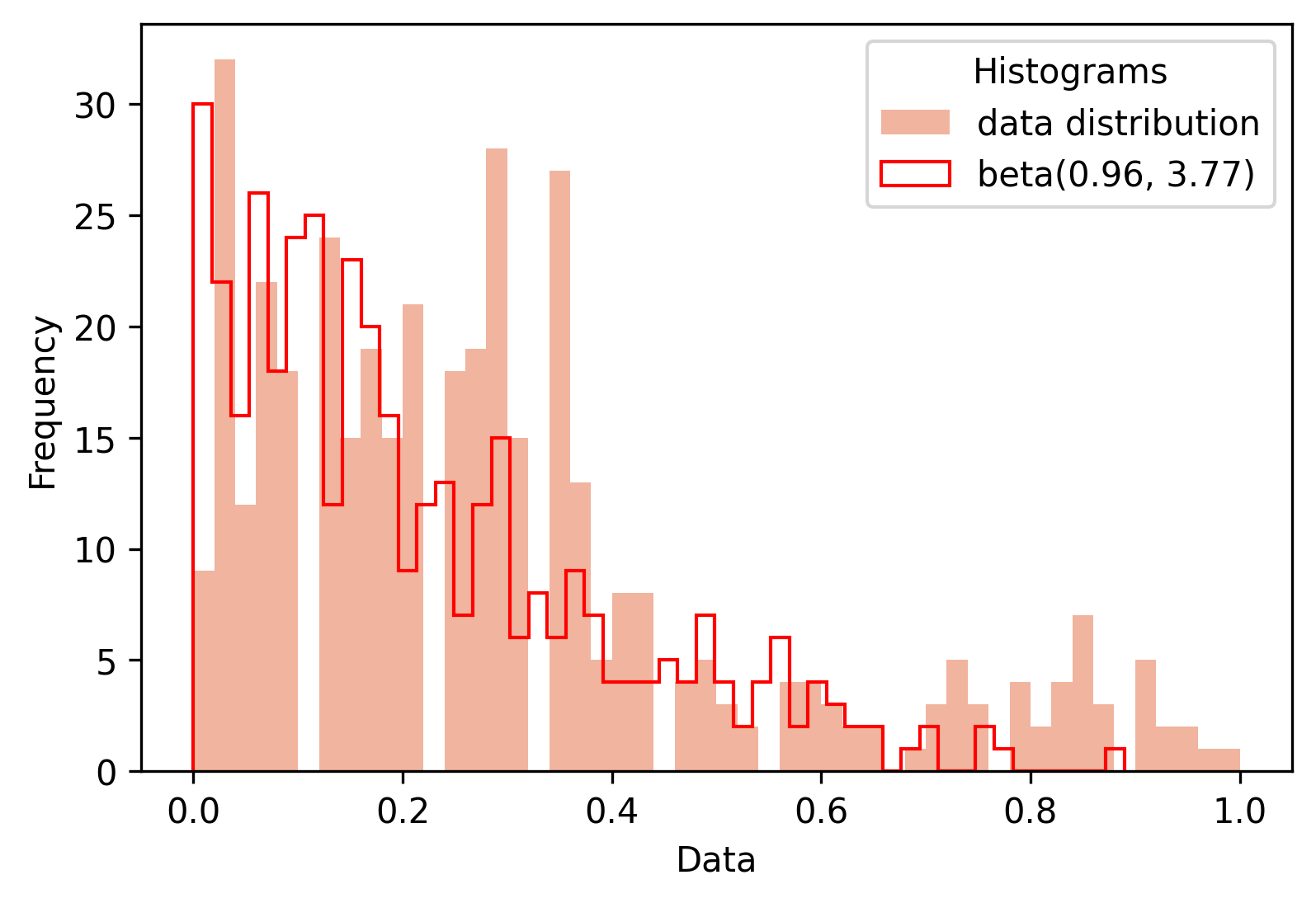}
\caption{Beta Distribution}
\end{subfigure}
\begin{subfigure}[b]{.3\textwidth}
\centering
\includegraphics[width=\textwidth]{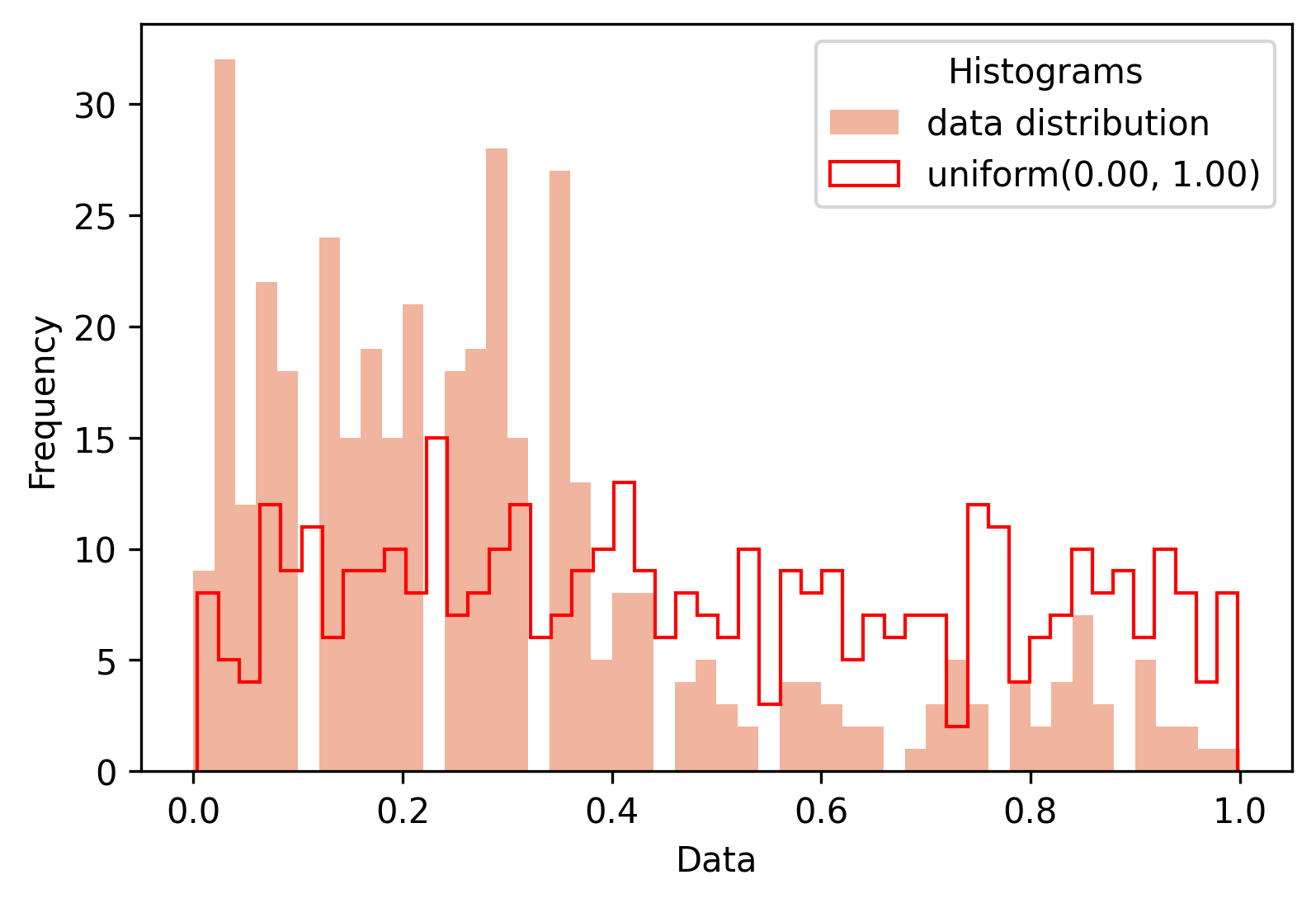}
\caption{Uniform Distribution}
\end{subfigure}
\begin{subfigure}[b]{.3\textwidth}
\centering
\includegraphics[width=\textwidth]{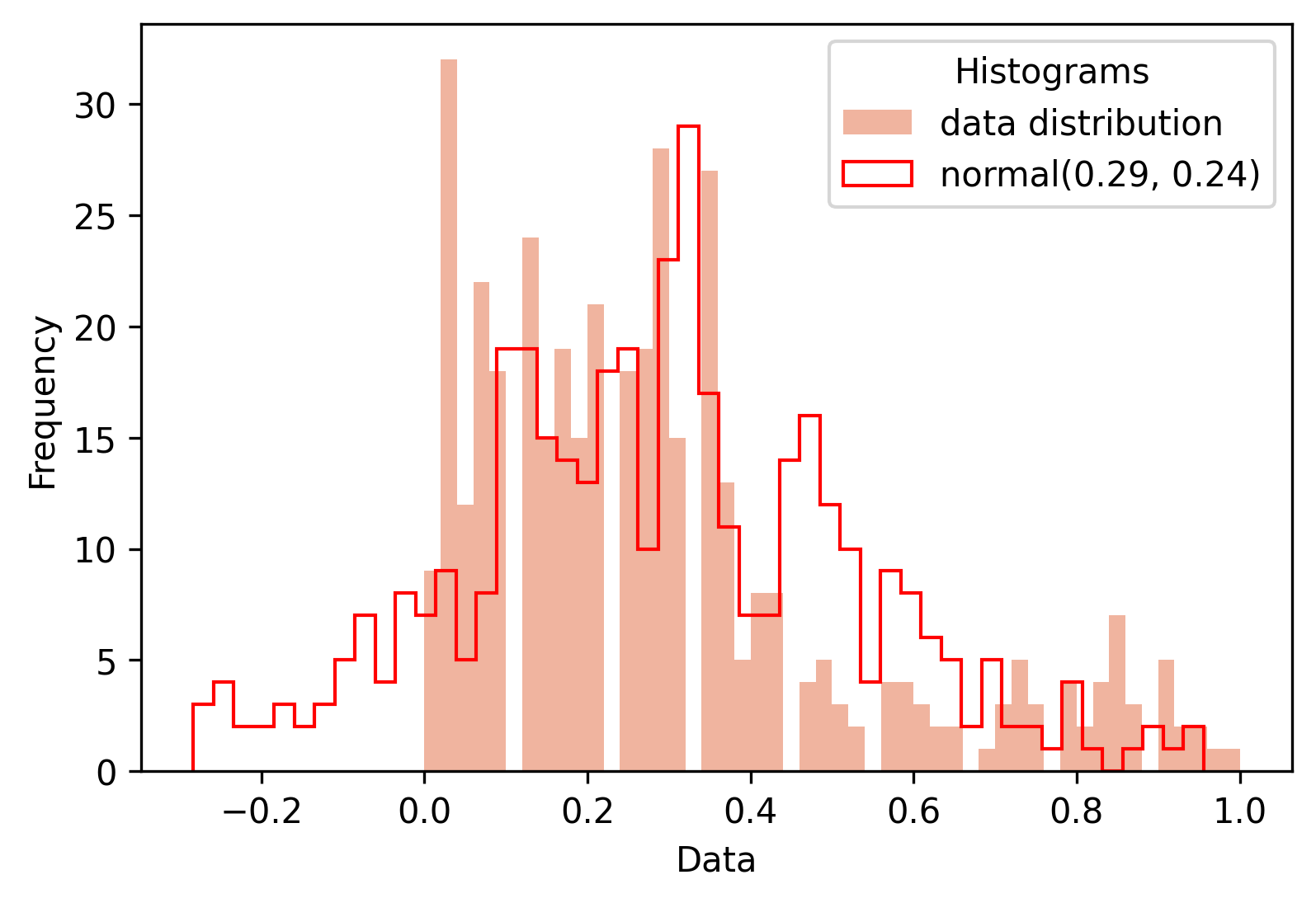}
\caption{Normal Distribution}
\end{subfigure}
\caption{The solid histogram shows the distribution of normalized Iris dataset. The hollow histogram(in red) shows the distribution of random samples of size equal to Iris drawn from parametrized distributions where the parameters have been estimated from the data.}
\label{fig:expbayes-distributions}
\end{figure*}

\section{Empirical Bayes}
The effectiveness of Beta distribution raises a question - is it possible to find the shape parameters in a principled manner? We analyze this question using insights from the Bayesian machine learning.

For any given dataset $\mathcal{D}$ and a model with parameters $\bm{\theta}$, the goal of Bayesian machine learning is to determine the posterior distribution $\pi(\bm{\theta} | \mathcal{D})$. A prior distribution over the parameters is given by $\pi(\bm{\theta})$ and the likelihood function is given as 
\[
L(\mathcal{D};\bm{\theta}) = \prod_{i=1}^{m} p(x_{i} | \bm{\theta}),
\]
where $x_{i} \in \mathcal{D}$. By the Bayes Law $\pi(\bm{\theta} | \mathcal{D}) \propto L(\mathcal{D};\bm{\theta}) \pi(\theta)$.

In a purely Bayesian model, we can set the prior $p(\bm{\theta})$ distribution to be a \emph{known} parametrized distribution. We can further assume a \emph{prior} on the hyperparameters of the parametrized distribution as well. However, doing this in practice is intractable for large dimensions or data points. A technique proposed in~\cite{bernardo2009bayesian, gelman1995bayesian} alleviates this problem by estimating the hyperpameters from the data using Maximum Likelihood Estimation (MLE) approach. These initial hyperparameters are then used to draw samples from the prior distribution. We use this insight to impose a known prior distribution over the parameters of the unitary gates and estimate the shape parameters directly from the data.

For the Beta distribution the MLE estimation of $\alpha$ and $\beta$ can be found as:

\begin{equation}
   \alpha_{MLE} \equiv  m \frac{\Gamma'(\alpha + \beta)}{\Gamma(\alpha+\beta)} - m \frac{\Gamma'(\alpha)}{\Gamma(\alpha)} + \sum_{i=1}^{m} x_{i} = 0
   \label{eq:alpha-mle}
\end{equation}

\begin{equation}
  \beta_{MLE}  \equiv m\frac{\Gamma'(\alpha + \beta)}{\Gamma(\alpha+\beta)} - m \frac{\Gamma'(\beta)}{\Gamma(\beta)} + \sum_{i=1}^{m} (1-x_{i}) = 0
  \label{eq:beta-mle}
\end{equation}




\begin{algorithm}
\caption{The \beinit Algorithm}
\label{alg:beinit}
\begin{algorithmic}[1]
\Procedure{\beinit}{$\mathcal{D},\eta, C(\theta), iters$}
    \State $\alpha, \beta \gets \Call{EB\_FIT}{\mathcal{D}}$
    \State $\mathcal{D}_{train}, \mathcal{D}_{test} \gets \Call{SPLIT}{\mathcal{D}}$
    \State $\theta_{i} \gets \Call{INIT}{\alpha, \beta}$
    \State $\theta \gets \theta_{i}$
    \For{$i= 0 \rightarrow iters$}
        \State $\sigma^{2}_{i} \gets \Call{Var}{\nabla_{\theta}L}$
        \State $\sigma^{2}_{n} \gets \dfrac{\eta}{(1 + i)^{\gamma + \sigma^{2}_{i}}}$
        \State {$\theta' \gets \theta + \mathcal{N}(0, \sigma^{2}_{n})$}

        \State $\theta^{i+1} \gets \Call{GRAD\_DESCENT}{\theta', \mathcal{D}_{train}, \eta, C(\theta)}$
    \EndFor

\EndProcedure
\end{algorithmic}
\end{algorithm}

\section{The \beinit Algorithm}
We propose an algorithm termed \beinit that is based on two key observations. First, drawing samples from a beta distribution for parametrizing the quantum circuit yields a much higher gradient variance for increasing number of qubits and second, it is possible to estimate the hyperparameters of the beta distribution from the data itself. The second observation is based on the empirical Bayes framework discussed earlier. 

Algorithm~\ref{alg:beinit} describes the overall procedure and is visually represented in Figure~\ref{sfig:beinit}. We take as input a  labelled dataset $\mathcal{D} = \{(x_{i}, y_{i})\}_{i=1}^{m}$, the learning rate $\eta$, a cost function $C(\theta)$ and the number of iteration steps \textit{iters} as a required input. In the first step, we estimate the $\alpha$ and $\beta$ parameters from all the available data using Equations~(\ref{eq:alpha-mle}) and (\ref{eq:beta-mle}). We then split this dataset into training and test subsets and initialize the circuit parameters from the discovered beta distribution parameters. 


During the training, we keep track of the gradient variance produced during an iteration $i$; $\sigma^{2}_i$. It has been shown that adding a perturbation during gradient descent can help a learning algorithm escape saddle points~\cite{jin2017escape}~\cite{ge2015escaping}. Inspired by a perturbation method proposed for deep learning~\cite{neelakantan2015adding}, we introduce noise in the parameter space by considering a normal distribution parameterized by 0 mean and $\sigma^{2}_{n}$ variance given by:
\begin{equation}
    \sigma^{2}_{n} = \frac{\eta}{(1 + i)^{(\gamma + \sigma^{2}_{i})}},
    \label{eq:normal-scaling}
\end{equation}
where $\eta$ and $\gamma$ are fixed parameters. One can interpret $\gamma$ as a constant additive bias that prevents a low noise perturbation (since $\sigma^{2}_{i}$ progressively decreases as the output of the circuit gets closer to the true label). Empirically, the choice of $\eta$ defines the amount of noise that will be generated per gradient descent step. We follow earlier works and choose $\eta \in \{0.01, 0.3, 1.0 \}$ and keep $\gamma = 0.55$. It must be noted that our method generates this perturbation in the \emph{parameter-space} as opposed the gradient space which is proposed by earlier related work~\cite{neelakantan2015adding}.

\section{Numerical Experiments}
In this section, we demonstrate numerical results that lend experimental evidence for our claims in the previous sections. The source code and data are available at \url{https://github.com/aicaffeinelife/BEINIT}

\begin{figure*}
\centering
\begin{subfigure}[b]{.45\textwidth}
\centering
\includegraphics[width=\textwidth]{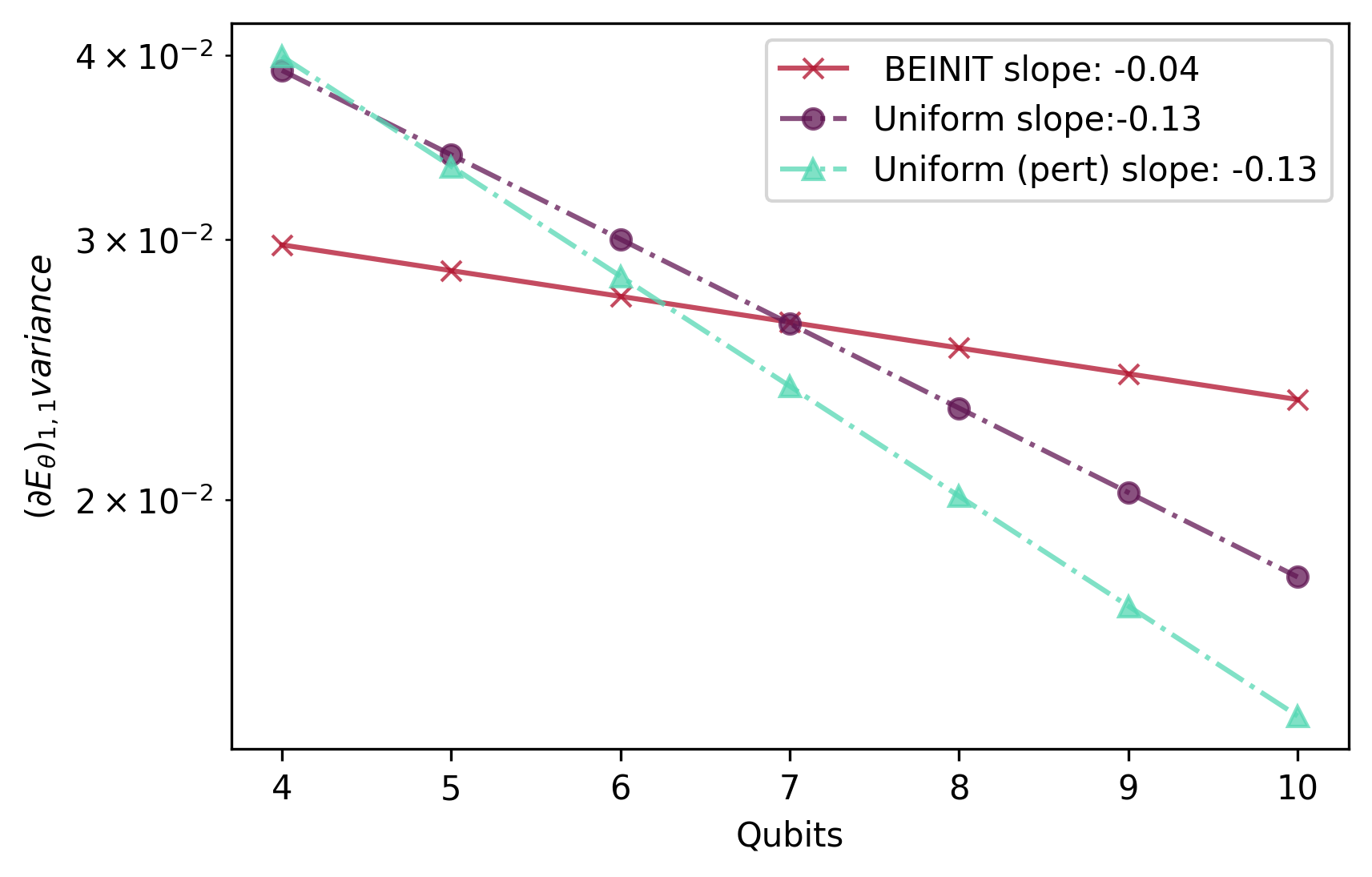}
\caption{Iris}
\end{subfigure}
\begin{subfigure}[b]{.45\textwidth}
\centering
\includegraphics[width=\textwidth]{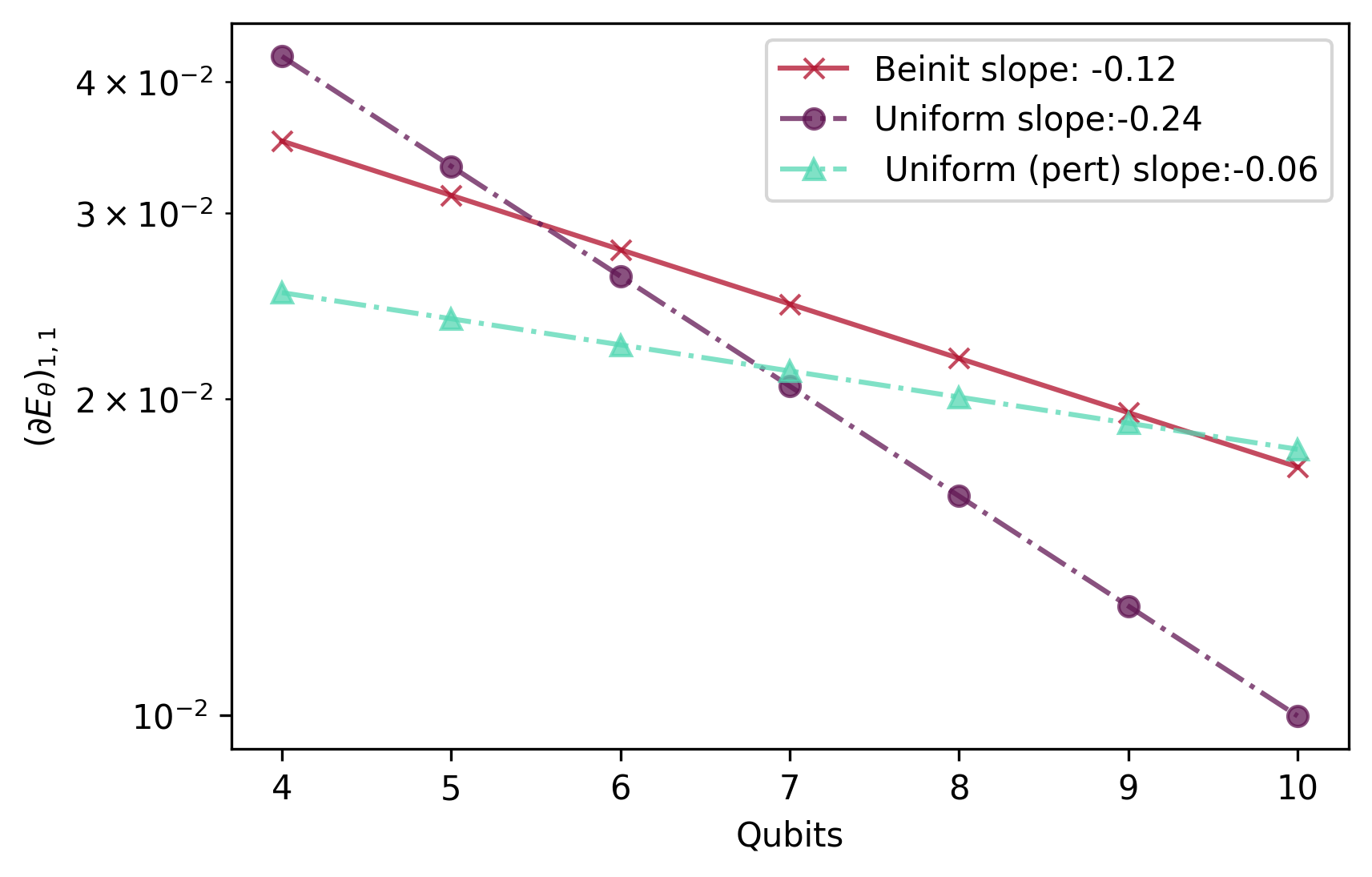}
\caption{Wine}
\end{subfigure}
\caption{The variance of the gradient of the first parameter is shown with increasing qubit sizes for three different initialization scenarios. Initialization with a beta distribution and added perturbation show the least degradation in variance with increasing qubit size.}

\label{fig:beinit_expt_results}
\end{figure*}

\subsection{Data Driven Initialization with Bayes Distribution}
In our exposition of the \beinit algorithm we had claimed that a data driven initialization process worked best with the Beta distribution. To verify this claim we extracted a two class subset from the original Iris dataset~\cite{dua2019}. We then normalized the data and estimated the parameters of three continuous distributions - Beta, Uniform and Normal using MLE estimation similar to Equations~(\ref{eq:alpha-mle}) and (\ref{eq:beta-mle}). The results of this experiment are shown in Figure~\ref{fig:expbayes-distributions}. 

The solid histogram is the normalized subset of the Iris dataset. The histogram contoured with red line is constructed by drawing an equivalent number of samples from the three continuous random distributions with their estimated parameters. We can see that for the case of uniform distribution, the estimated upper and lower bounds correspond to the minimum and maximum value of the data and the random distribution is evenly spread around those bounds. A similar observation can be made about the case of the normal distribution where the random distribution concentrates heavily around the mean of the underlying data. In contrast to these two cases, the Beta distribution closely conforms to the contours of the data. These results indicate that the Beta distribution is a much more data-sensitive distribution and can help provide a parameter initialization such that the parameters do not initially concentrate around the first or second order moments of the underlying data.

\subsection{Experiments with Increasing Number of Qubits}
\begin{figure}[h!]
    \centering
    \includegraphics[width=.95\columnwidth]{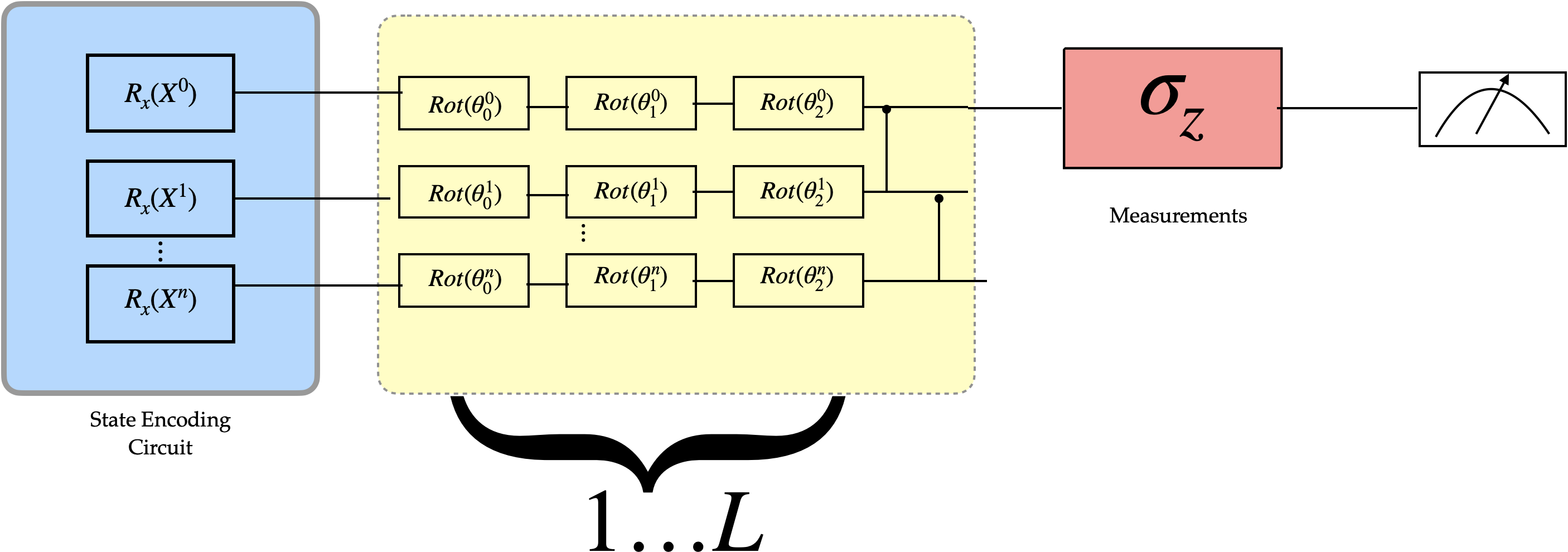}
    \caption{A graphical representation of the circuit used in our experiments}
    \label{fig:expt_circuit}
\end{figure}

We demonstrate the effectiveness of our algorithm on the Iris and Wine datasets~\cite{dua2019}. To study the effect of the \beinit algorithm on the gradient variance with increasing number of qubits, we compare against two baseline cases. In the first case, we initialize the QNN's parameters using the uniform distribution as per~\cite{mcclean2018barren} with the parameters estimated from the normalized data distribution. In the second case, we keep the same initialization as before, but introduce the perturbation as in Equation~(\ref{eq:normal-scaling}).  Both datasets considered in this study are multi-class classification problems, but we convert them into a binary classification problem by taking only first two classes. The labels are then binarized to $\pm 1$. In the case of the Wine dataset, we reduce the dimensionality of the data by performing a principal component analysis (PCA) such that $d = 2$, where $d$ indicates the dimensionality of a given data vector. We choose the number of qubits $q \in \{4, 5, 6, 7, 8, 9, 10 \}$. All our experiments are performed on the default quantum simulator provided by the Pennylane~\cite{bergholm2018pennylane} library and we use the Nesterov momentum optimizer~\cite{nesterov1983method} for all experiments.\\

\noindent
The QNN architecture used in our studies is shown in Figure~\ref{fig:expt_circuit}. The first stage in our architecture is a standard state encoding circuit that maps $\vec{x} \in \mathbb{R}^{d} \mapsto \ket{\psi(x)} \in \mathbb{C}^{q}$. For each dimension $j = 0\dots d; d \in \{2, 4\}$,  we encode $x_{j} \in \vec{x}$ as:
\begin{equation}
    \ket{\psi(x)}^{(j)} = e^{-i x_{j} \sigma_{x}},
    \label{eq:angle-encoding}
\end{equation}
where $\ket{\psi(x)}^{(j)}$ indicates the $j$-th component of the resulting input quantum state and the RHS is the X rotation gate. Since we wanted to study the effect of increasing number of qubits, we kept the data encoding process as simple as possible. 

The second stage of our architecture comprises of a cascade of parametrized rotation gates $R(\theta)$ arranged in a $q \times 3$ shape i.e. each component of the input state is passed through a succession of three rotation gates. Their outputs are then entangled using CNOT gates. This block is then arranged in $L$ layers. We choose a single Pauli Z operator applied on the first qubit as the objective operator for the output state $\ket{\psi_{o}}$.

Figure~\ref{fig:beinit_expt_results} shows an interpolated line computed from the raw gradient values for different cases. We can observe that for both datasets, initializing with a beta distribution proves to be beneficial in reducing the likelihood of a 2-design from forming. We can also see that in certain cases (e.g. Wine dataset), perturbing the gradient even with uniform distribution can have a beneficial effect on the variance of the gradient. It is interesting to note that in both datasets, a variational quantum circuit initialized with uniform distribution exhibits a similar decay in variance as with a random circuit of~\cite{mcclean2018barren}. This strongly indicates that a VQCs initialized with a uniform distribution have a higher likelihood of exhibiting a unitary 2-design.

\subsection{Experiments with Deeper Circuits}
\begin{figure*}
\centering
\begin{subfigure}[b]{.45\textwidth}
\centering
\includegraphics[width=\textwidth]{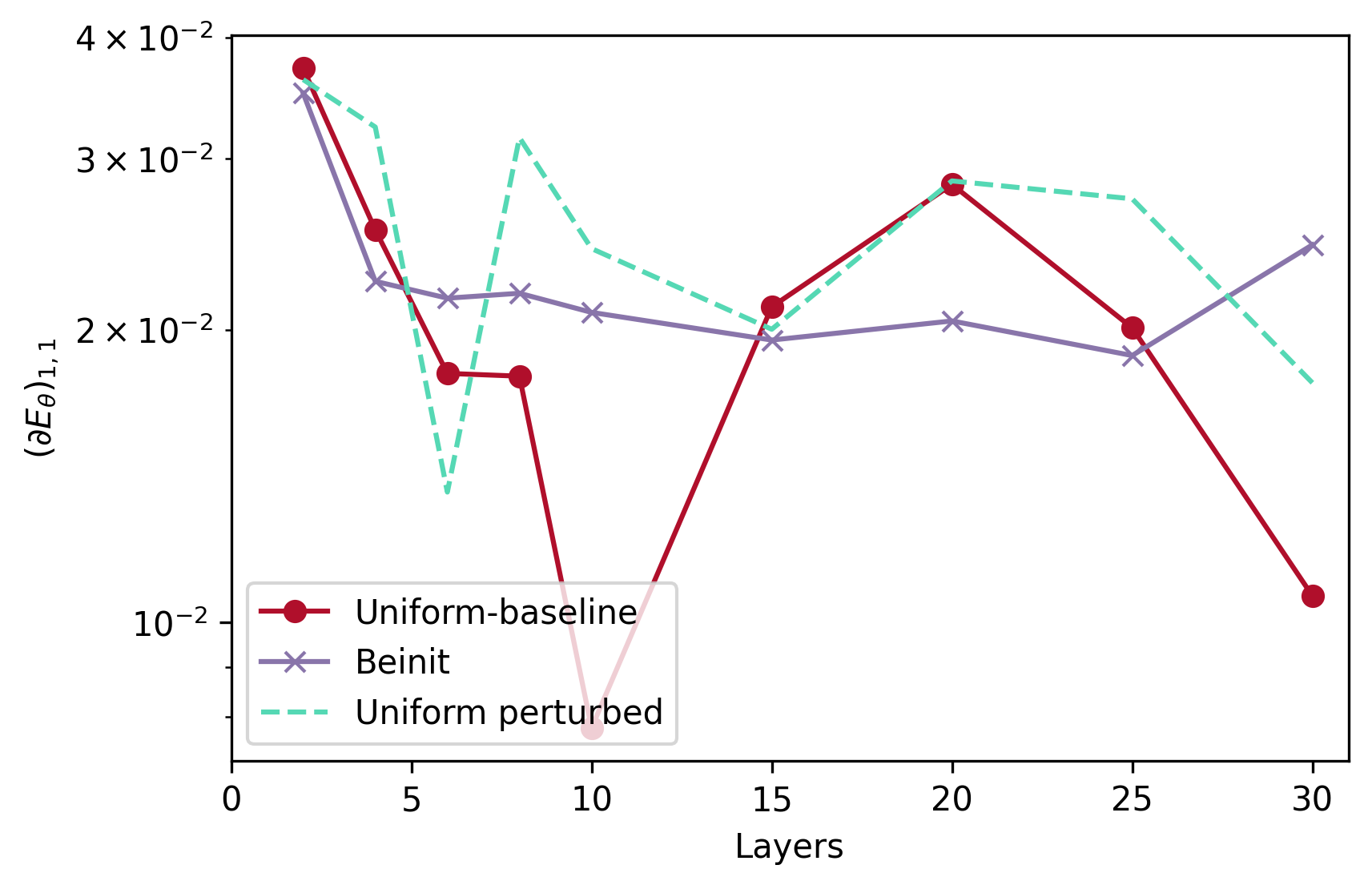}
\caption{Iris}
\end{subfigure}
\begin{subfigure}[b]{.45\textwidth}
\centering
\includegraphics[width=\textwidth]{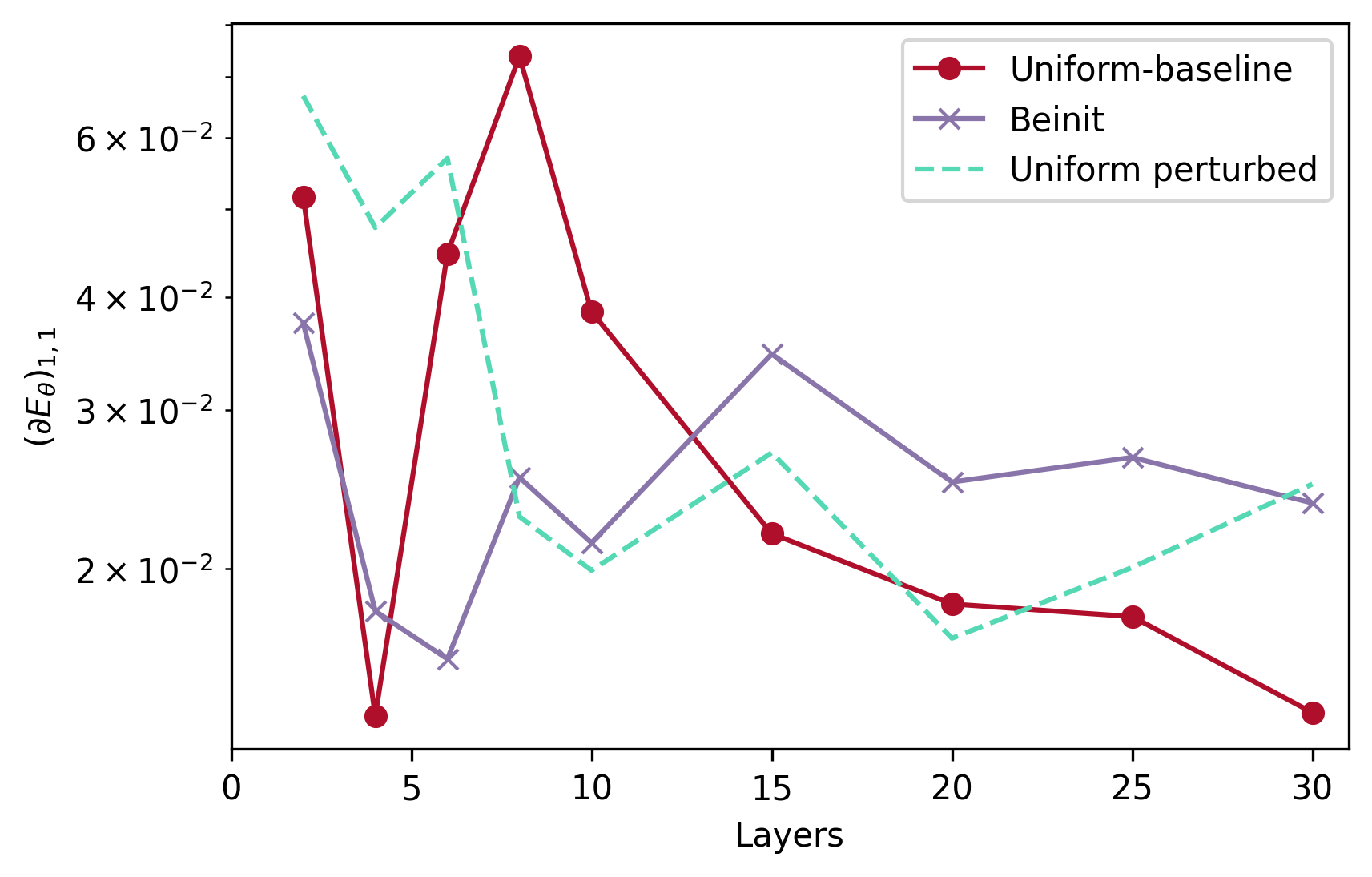}
\caption{Wine}
\end{subfigure}
\caption{The variance of gradient of first parameter is shown with increasing number of layers. The uniform distribution with no perturbation shows a downward trend with increasing layers. Beta and Uniform initialization with parameter perturbation result in a much more stable variance with beta initialization outperforming the uniform case.}
\label{fig:beinit_layer_expt_results}
\end{figure*}

We also performed experiments with QNNs that were several layers deep on the two datasets above. In this set of experiments, we kept the number of qubits to be constant at 4 and varied the number of layers $L \in \{2, 4, 6, 8, 10, 15, 20, 25, 30 \}$. Besides these changes, we kept the experimental setup similar to the qubit experiments before.

The variance of gradient of the first parameters in different cases is shown in Figure~\ref{fig:beinit_layer_expt_results} shows the result of our experiments. As expected, the uniform distribution with no perturbation shows a poor scaling with increasing number of layers. With parameter perturbation, both uniform and beta initialization show a more stable behavior. In the case of the Iris dataset, the \beinit procedure results in a steadier variance as compared to the uniform initialization with perturbation. The performance of \beinit is also evidenced in the case of the Wine dataset. The results of these experiments shed light on the efficacy of both perturbation and data driven beta initialization for deep QNN circuits.





\section{Discussion}

In this paper, two crucial questions regarding solving the barren plateau problem have been raised. We first explored different probability distributions and found that the choice of probability distribution for drawing the initial set of parameters has a direct influence on the variance of the gradient with complex circuits (i.e., circuits with a large number of qubits).  Our experiments show that the Beta distribution is a highly data-sensitive distribution and initializing from it can lead to much lower degradation in gradient variance. This result is general and can be applied to initialization of parameters in different VQA algorithms. In the case of QNNs, the data guides the end result of optimization. We thus design a data-driven initialization process that incorporates the insights from data (e.g., determining the hyperparameters of the initializing distribution) to guide the overall initialization process. The observations from our experiments confirm that this data-driven initialization process can have a strong positive influence on the overall training process and reduce the chances of a barren plateaus. Our experimental insights lead us to develop a conjecture that explores the relationship between barren plateaus in optimization and the characteristics of the initalizing distribution. To the best of our knowledge, this is the first work that considers such a relationship in the context of VQA.



Our second question opened a novel direction in the exploration of solutions to avoid the barren plateau problem. In the previous works, no attention has been paid to the effect of \emph{perturbation} during the course of training and its subsequent role in improving the variance of the circuit. We devised the \beinit procedure to combine our earlier insights along with a parameter perturbation scheme and showed that it is effective in improving the variance of the circuit with increasing number of qubits. Our experiments with the initialization and perturbation are shedding light on the barren plateau problem and suggest that a solution to overcome it may lie in finding a good data driven initialization and perturbation during the course of gradient descent. Another promising research direction is to investigate how the proposed technique affects another variational quantum approach, namely, the quantum approximate optimization algorithm (QAOA). Its acceleration at increasing circuit depth and number of qubits including techniques to deal with the barren plateau is a subject of major efforts \cite{shaydulin2019multistart,shaydulin2021classical}.

\section*{Acknowledgments}

This work was supported in part with funding from the ONISQ program of the Defense Advanced Research Projects Agency (DARPA). The views, opinions and/or findings expressed are those of the authors and should not be interpreted as representing the official views or policies of the Department of Defense or the U.S. Government.

\bibliographystyle{IEEEtran}
\bibliography{ref.bib}

\end{document}